\PassOptionsToPackage{unicode}{hyperref}
\PassOptionsToPackage{hyphens}{url}
\PassOptionsToPackage{dvipsnames,svgnames,x11names}{xcolor}
\documentclass[
]{article}
\usepackage{amsmath,amssymb}
\usepackage{lmodern}
\usepackage{iftex}
\ifPDFTeX
  \usepackage[T1]{fontenc}
  \usepackage[utf8]{inputenc}
  \usepackage{textcomp} 
\else 
  \usepackage{unicode-math}
  \defaultfontfeatures{Scale=MatchLowercase}
  \defaultfontfeatures[\rmfamily]{Ligatures=TeX,Scale=1}
\fi
\IfFileExists{upquote.sty}{\usepackage{upquote}}{}
\IfFileExists{microtype.sty}{
  \usepackage[]{microtype}
  \UseMicrotypeSet[protrusion]{basicmath} 
}{}
\makeatletter
\@ifundefined{KOMAClassName}{
  \IfFileExists{parskip.sty}{%
    \usepackage{parskip}
  }{
    \setlength{\parindent}{0pt}
    \setlength{\parskip}{6pt plus 2pt minus 1pt}}
}{
  \KOMAoptions{parskip=half}}
\makeatother
\usepackage{xcolor}
\usepackage{graphicx}
\makeatletter
\def\maxwidth{\ifdim\Gin@nat@width>\linewidth\linewidth\else\Gin@nat@width\fi}
\def\maxheight{\ifdim\Gin@nat@height>\textheight\textheight\else\Gin@nat@height\fi}
\makeatother
\setkeys{Gin}{width=\maxwidth,height=\maxheight,keepaspectratio}
\makeatletter
\def\fps@figure{htbp}
\makeatother
\providecommand{\tightlist}{%
  \setlength{\itemsep}{0pt}\setlength{\parskip}{0pt}}
\NewDocumentCommand\citeproctext{}{}
\NewDocumentCommand\citeproc{mm}{%
  \begingroup\def\citeproctext{#2}\cite{#1}\endgroup}
\makeatletter
 \let\@cite@ofmt\@firstofone
 \def\@biblabel#1{}
 \def\@cite#1#2{{#1\if@tempswa , #2\fi}}
\makeatother
\newlength{\cslhangindent}
\newlength{\csllabelwidth}
\newenvironment{CSLReferences}[2] 
 {\begin{list}{}{%
  \setlength{\itemindent}{0pt}
  \setlength{\leftmargin}{0pt}
  \setlength{\parsep}{0pt}
  \ifodd #1
   \setlength{\leftmargin}{\cslhangindent}
   \setlength{\itemindent}{-1\cslhangindent}
  \fi
  \setlength{\itemsep}{#2\baselineskip}}}
 {\end{list}}
\usepackage{calc}

\ifLuaTeX
\usepackage[bidi=basic]{babel}
\else
\usepackage[bidi=default]{babel}
\fi
\babelprovide[main,import]{american}

\def\languageshorthands#1{}
\ifLuaTeX
  \usepackage{selnolig}  
\fi
\IfFileExists{bookmark.sty}{\usepackage{bookmark}}{\usepackage{hyperref}}
\IfFileExists{xurl.sty}{\usepackage{xurl}}{} 
\hypersetup{
  pdftitle={pythonradex: a fast Python re-implementation of RADEX with
extended functionality},
  pdfauthor={Gianni Cataldi},
  pdflang={en-US},
  colorlinks=true,
  linkcolor={Maroon},
  filecolor={Maroon},
  citecolor={Blue},
  urlcolor={Blue},
  pdfcreator={LaTeX via pandoc}}

\title{pythonradex: a fast Python re-implementation of RADEX with
extended functionality}

\definecolor{c53baa1}{RGB}{83,186,161}
\definecolor{c202826}{RGB}{32,40,38}
\def \rorglobalscale {0.1}
\newcommand{\rorlogo}{%
\begin{tikzpicture}[y=1cm, x=1cm, yscale=\rorglobalscale,xscale=\rorglobalscale, every node/.append style={scale=\rorglobalscale}, inner sep=0pt, outer sep=0pt]
  \begin{scope}[even odd rule,line join=round,miter limit=2.0,shift={(-0.025, 0.0216)}]
    \path[fill=c53baa1,nonzero rule,line join=round,miter limit=2.0] (1.8164, 3.012) -- (1.4954, 2.5204) -- (1.1742, 3.012) -- (1.8164, 3.012) -- cycle;
    \path[fill=c53baa1,nonzero rule,line join=round,miter limit=2.0] (3.1594, 3.012) -- (2.8385, 2.5204) -- (2.5172, 3.012) -- (3.1594, 3.012) -- cycle;
    \path[fill=c53baa1,nonzero rule,line join=round,miter limit=2.0] (1.1742, 0.0669) -- (1.4954, 0.5588) -- (1.8164, 0.0669) -- (1.1742, 0.0669) -- cycle;
    \path[fill=c53baa1,nonzero rule,line join=round,miter limit=2.0] (2.5172, 0.0669) -- (2.8385, 0.5588) -- (3.1594, 0.0669) -- (2.5172, 0.0669) -- cycle;
    \path[fill=c202826,nonzero rule,line join=round,miter limit=2.0] (3.8505, 1.4364).. controls (3.9643, 1.4576) and (4.0508, 1.5081) .. (4.1098, 1.5878).. controls (4.169, 1.6674) and (4.1984, 1.7642) .. (4.1984, 1.8777).. controls (4.1984, 1.9719) and (4.182, 2.0503) .. (4.1495, 2.1132).. controls (4.1169, 2.1762) and (4.0727, 2.2262) .. (4.0174, 2.2635).. controls (3.9621, 2.3006) and (3.8976, 2.3273) .. (3.824, 2.3432).. controls (3.7505, 2.359) and (3.6727, 2.367) .. (3.5909, 2.367) -- (2.9676, 2.367) -- (2.9676, 1.8688).. controls (2.9625, 1.8833) and (2.9572, 1.8976) .. (2.9514, 1.9119).. controls (2.9083, 2.0164) and (2.848, 2.1056) .. (2.7705, 2.1791).. controls (2.6929, 2.2527) and (2.6014, 2.3093) .. (2.495, 2.3487).. controls (2.3889, 2.3881) and (2.2728, 2.408) .. (2.1468, 2.408).. controls (2.0209, 2.408) and (1.905, 2.3881) .. (1.7986, 2.3487).. controls (1.6925, 2.3093) and (1.6007, 2.2527) .. (1.5232, 2.1791).. controls (1.4539, 2.1132) and (1.3983, 2.0346) .. (1.3565, 1.9436).. controls (1.3504, 2.009) and (1.3351, 2.0656) .. (1.3105, 2.1132).. controls (1.2779, 2.1762) and (1.2338, 2.2262) .. (1.1785, 2.2635).. controls (1.1232, 2.3006) and (1.0586, 2.3273) .. (0.985, 2.3432).. controls (0.9115, 2.359) and (0.8337, 2.367) .. (0.7519, 2.367) -- (0.1289, 2.367) -- (0.1289, 0.7562) -- (0.4837, 0.7562) -- (0.4837, 1.4002) -- (0.6588, 1.4002) -- (0.9956, 0.7562) -- (1.4211, 0.7562) -- (1.0118, 1.4364).. controls (1.1255, 1.4576) and (1.2121, 1.5081) .. (1.2711, 1.5878).. controls (1.2737, 1.5915) and (1.2761, 1.5954) .. (1.2787, 1.5991).. controls (1.2782, 1.5867) and (1.2779, 1.5743) .. (1.2779, 1.5616).. controls (1.2779, 1.4327) and (1.2996, 1.3158) .. (1.3428, 1.2113).. controls (1.3859, 1.1068) and (1.4462, 1.0176) .. (1.5237, 0.944).. controls (1.601, 0.8705) and (1.6928, 0.8139) .. (1.7992, 0.7744).. controls (1.9053, 0.735) and (2.0214, 0.7152) .. (2.1474, 0.7152).. controls (2.2733, 0.7152) and (2.3892, 0.735) .. (2.4956, 0.7744).. controls (2.6016, 0.8139) and (2.6935, 0.8705) .. (2.771, 0.944).. controls (2.8482, 1.0176) and (2.9086, 1.1068) .. (2.952, 1.2113).. controls (2.9578, 1.2253) and (2.9631, 1.2398) .. (2.9681, 1.2544) -- (2.9681, 0.7562) -- (3.3229, 0.7562) -- (3.3229, 1.4002) -- (3.4981, 1.4002) -- (3.8349, 0.7562) -- (4.2603, 0.7562) -- (3.8505, 1.4364) -- cycle(0.9628, 1.7777).. controls (0.9438, 1.7534) and (0.92, 1.7357) .. (0.8911, 1.7243).. controls (0.8623, 1.7129) and (0.83, 1.706) .. (0.7945, 1.7039).. controls (0.7588, 1.7015) and (0.7252, 1.7005) .. (0.6932, 1.7005) -- (0.4839, 1.7005) -- (0.4839, 2.0667) -- (0.716, 2.0667).. controls (0.7477, 2.0667) and (0.7805, 2.0643) .. (0.8139, 2.0598).. controls (0.8472, 2.0553) and (0.8768, 2.0466) .. (0.9025, 2.0336).. controls (0.9282, 2.0206) and (0.9496, 2.0021) .. (0.9663, 1.9778).. controls (0.9829, 1.9534) and (0.9914, 1.9209) .. (0.9914, 1.8799).. controls (0.9914, 1.8362) and (0.9819, 1.8021) .. (0.9628, 1.7777) -- cycle(2.6125, 1.3533).. controls (2.5889, 1.2904) and (2.5553, 1.2359) .. (2.5112, 1.1896).. controls (2.4672, 1.1433) and (2.4146, 1.1073) .. (2.3529, 1.0814).. controls (2.2916, 1.0554) and (2.2228, 1.0427) .. (2.1471, 1.0427).. controls (2.0712, 1.0427) and (2.0026, 1.0557) .. (1.9412, 1.0814).. controls (1.8799, 1.107) and (1.8272, 1.1433) .. (1.783, 1.1896).. controls (1.7391, 1.2359) and (1.7052, 1.2904) .. (1.6817, 1.3533).. controls (1.6581, 1.4163) and (1.6465, 1.4856) .. (1.6465, 1.5616).. controls (1.6465, 1.6359) and (1.6581, 1.705) .. (1.6817, 1.7687).. controls (1.7052, 1.8325) and (1.7388, 1.8873) .. (1.783, 1.9336).. controls (1.8269, 1.9799) and (1.8796, 2.0159) .. (1.9412, 2.0418).. controls (2.0026, 2.0675) and (2.0712, 2.0804) .. (2.1471, 2.0804).. controls (2.223, 2.0804) and (2.2916, 2.0675) .. (2.3529, 2.0418).. controls (2.4143, 2.0161) and (2.467, 1.9799) .. (2.5112, 1.9336).. controls (2.5551, 1.8873) and (2.5889, 1.8322) .. (2.6125, 1.7687).. controls (2.636, 1.705) and (2.6477, 1.6359) .. (2.6477, 1.5616).. controls (2.6477, 1.4856) and (2.636, 1.4163) .. (2.6125, 1.3533) -- cycle(3.8015, 1.7777).. controls (3.7825, 1.7534) and (3.7587, 1.7357) .. (3.7298, 1.7243).. controls (3.701, 1.7129) and (3.6687, 1.706) .. (3.6333, 1.7039).. controls (3.5975, 1.7015) and (3.5639, 1.7005) .. (3.5319, 1.7005) -- (3.3226, 1.7005) -- (3.3226, 2.0667) -- (3.5547, 2.0667).. controls (3.5864, 2.0667) and (3.6192, 2.0643) .. (3.6526, 2.0598).. controls (3.6859, 2.0553) and (3.7155, 2.0466) .. (3.7412, 2.0336).. controls (3.7669, 2.0206) and (3.7883, 2.0021) .. (3.805, 1.9778).. controls (3.8216, 1.9534) and (3.8301, 1.9209) .. (3.8301, 1.8799).. controls (3.8301, 1.8362) and (3.8206, 1.8021) .. (3.8015, 1.7777) -- cycle;
  \end{scope}
\end{tikzpicture}
}

\usepackage[affil-it]{authblk}
\usepackage{orcidlink}
\author[1]{Gianni Cataldi%
    \,\orcidlink{0000-0002-2700-9676}\,%
    }

\affil[1]{National Astronomical Observatory of Japan, 2-21-1 Osawa,
Mitaka, Tokyo 181-8588, Japan%
    \,\protect\href{https://ror.org/052rrw050}{\protect\rorlogo}\,%
  }

\date{25 February 2026}
\begin{document}
\maketitle

\section{Summary}\label{summary}

A common task in astronomical research is to estimate the physical
parameters (temperature, mass, density etc.) of a gas by using observed
line emission. This often requires a calculation of how the radiation
propagates via emission and absorption (``radiative transfer''). In
radio and infrared astronomy, the Fortran code \texttt{RADEX}
(\citeproc{ref-vanderTak2007}{van der Tak et al., 2007}) is a popular
tool to solve the non-LTE radiative transfer of a uniform medium in a
simplified geometry. This paper presents \texttt{pythonradex}, a Python re-implementation of
\texttt{RADEX}. Written in Python, it provides an easy and intuitive user interface, improved performance as well as additional functionality not included in \texttt{RADEX}, such as continuum effects and overlapping lines. In addition, \texttt{pythonradex} provides a self-consistent computation of the total flux for all geometries, including spherical geometries. 

\section{Statement of need}\label{statement-of-need}

Modern astronomical facilities such as the Atacama Large
Millimeter/submillimeter Array (ALMA) or the James Webb Space Telescope
(JWST) are providing a wealth of line emission data at radio and
infrared wavelengths. These data are crucial to constrain the physical
and chemical properties of various astrophysical environments.

To interpret such data, a radiative transfer calculation is typically
used (see \citeproc{ref-Rybicki1985}{Rybicki \& Lightman, 1985}, for an
introduction to radiative transfer). For a given set of input parameters
describing the source (temperature, density, geometry, etc.), one
calculates the amount of radiation reaching the telescope. The input
parameters are then adjusted such that the predicted flux matches the
observations.

If the medium is dense enough, local thermodynamic equilibrium (LTE) can
be assumed. This considerably simplifies the radiative transfer
calculation. A non-LTE calculation is considerably more complex and
computationally expensive because the fractional population of the
molecular energy levels needs to be solved for numerically (e.g.
\citeproc{ref-vanderTak2007}{van der Tak et al., 2007}). Typically, an
iterative approach is used: from a first guess of the level populations,
the radiation field is calculated. This radiation field is then used to
solve for updated level populations. The iterations continue until
convergence is reached.

Various codes are available to solve the radiative transfer. Codes
solving the radiative transfer in 3D are used for detailed calculations
of sources with well-known geometries. Examples include
\texttt{RADMC-3D} (\citeproc{ref-Dullemond2012}{Dullemond et al., 2012})
and \texttt{LIME} (\citeproc{ref-Brinch2010}{Brinch \& Hogerheijde,
2010}). However, a full 3D calculation is often too computationally
expensive if a large parameter space needs to be explored, in particular
in non-LTE. 1D codes that quickly provide an approximate solution are a
commonly used alternative. In this respect, the 1D non-LTE code
\texttt{RADEX} (\citeproc{ref-vanderTak2007}{van der Tak et al., 2007})
has gained considerable popularity: as of February 10, 2026, the paper
presenting \texttt{RADEX} (\citeproc{ref-vanderTak2007}{van der Tak et
al., 2007}) has 1463 citations. The Fortran code \texttt{RADEX} solves
the radiative transfer of a uniform medium using an escape probability
formalism.

The Python programming language is now widely used in astronomy. Still,
no Python version of \texttt{RADEX} is available, although some Python
wrappers such as \texttt{SpectralRadex}
(\citeproc{ref-SpectralRadex}{Holdship et al., 2023}) or
\texttt{ndradex} (\citeproc{ref-ndradex}{Taniguchi, 2019}), and even a
Julia version (\texttt{Jadex}, \citeproc{ref-jadex}{Svoboda, 2022}),
exist. Furthermore, \texttt{RADEX} cannot take into account the effects
of an internal continuum field (typically arising from dust that is
mixed with the gas), nor cross-excitation effects arising when
transitions overlap in frequency. The \texttt{pythonradex} code
addresses these concerns.

\section{Implementation}\label{implementation}

\texttt{pythonradex} is written in Python and implements the accelerated
lambda iteration (ALI) scheme presented by Rybicki \& Hummer
(\citeproc{ref-Rybicki1992}{1992}). Like \texttt{RADEX}, an escape
probability equation is used to calculate the radiation field for a
given level population. This allows solving the radiative transfer
iteratively. To speed up the convergence, Ng acceleration
(\citeproc{ref-Ng1974}{Ng, 1974}) is employed. To improve performance,
critical parts of the code are just-in-time compiled using Numba
(\citeproc{ref-Lam2015}{Lam et al., 2015}).

\texttt{pythonradex} supports four geometries: two static geometries
(slab and sphere; note that \texttt{RADEX} does not support a static slab), and two large-velocity-gradient (LVG) geometries
(again slab and sphere). In the LVG approximation, it is assumed that
all regions of the source are Doppler shifted with respect to each other
due to a velocity gradient. This means that all photons escape the
source, unless absorbed locally (i.e.~close to the emission location,
e.g. \citeproc{ref-Elitzur1992}{Elitzur, 1992}).

Currently, effects of internal continuum and overlapping lines can only
be included for the static geometries. Another limitation is that only a
single molecule can be considered at a time. Thus, solving the radiative
transfer of overlapping lines of different molecules is not supported
yet. Also, treating overlapping lines adds considerable computational
cost because averages over line profiles need to be calculated.

Like \texttt{RADEX}, \texttt{pythonradex} needs a file in LAMDA-format
as input to read the molecular data. Such files can for example be
downloaded from the LAMDA (\citeproc{ref-Schoier2005}{Schöier et al.,
2005}) or EMAA (\citeproc{ref-EMAA}{\emph{EMAA}, 2021}) databases.

\section{Benchmarking}\label{benchmarking}

\begin{figure}
\centering
\includegraphics[keepaspectratio]{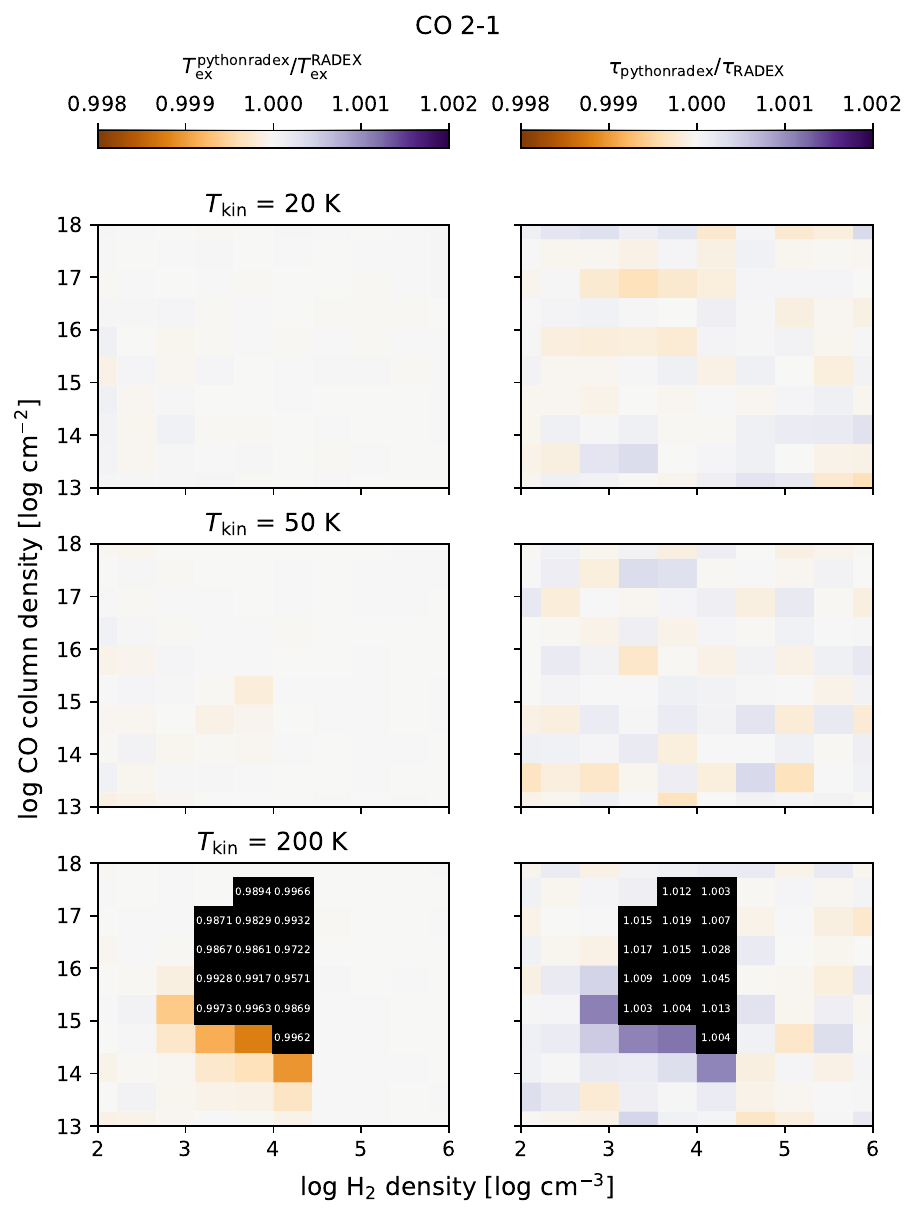}
\caption{The ratio of CO 2-1 excitation temperature (left column) and
optical depth (right column) computed with \texttt{pythonradex} and
\texttt{RADEX} for a static sphere. Each panel shows a parameter space
of H\(_2\) density and column density for a fixed kinetic temperature.
Values exceeding the colorbar range are shown in black with the
corresponding value in white text.\label{fig:pythonradex_vs_radex}}
\end{figure}

\begin{figure}
\centering
\includegraphics[keepaspectratio]{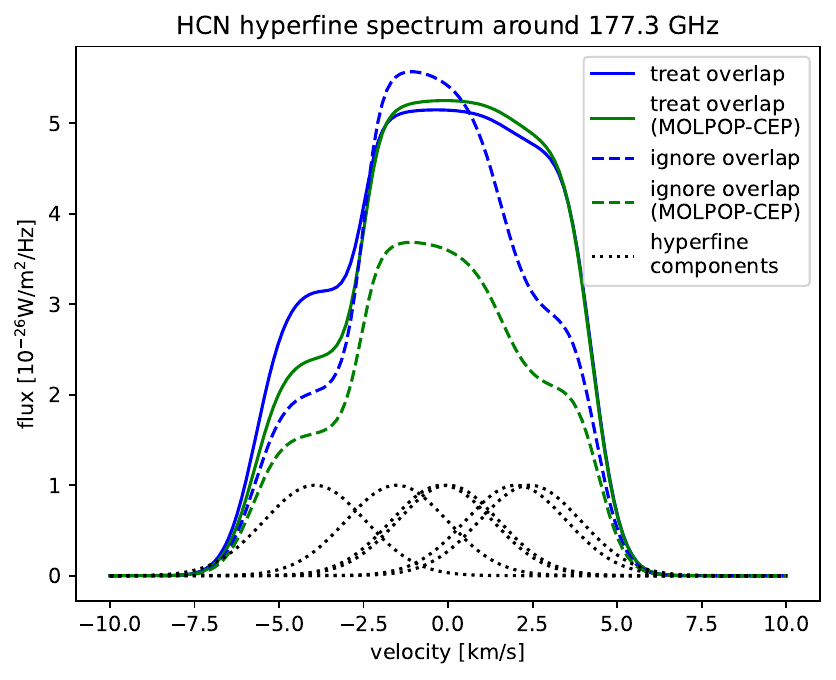}
\caption{Spectrum of HCN around 177.3 GHz computed with
\texttt{pythonradex} and \texttt{MOLPOP-CEP} for a static slab. Good
agreement is found when treating line overlap. Interestingly, the
spectra differ somewhat when ignoring overlap. The positions and widths
of the individual hyperfine components are illustrated by the black
dotted lines.\label{fig:HCN_spectrum}}
\end{figure}

\texttt{pythonradex} was benchmarked against \texttt{RADEX} for a number
of example problems, generally with excellent agreement (see
\autoref{fig:pythonradex_vs_radex} for an example). To test the
treatment of overlapping lines, \texttt{pythonradex} was tested against
the \texttt{MOLPOP-CEP} code (\citeproc{ref-AsensioRamos2018}{Asensio
Ramos \& Elitzur, 2018}), again showing good agreement, as illustrated
in \autoref{fig:HCN_spectrum}.

\section{Performance advantage}\label{performance-advantage}

Both \texttt{pythonradex} and \texttt{RADEX} are single-threaded. To
compare their performance, we consider the calculation of a grid of
models over a parameter space spanning 20 values in each of kinetic
temperature, column density and H\(_2\) density (i.e.~a total of 8000
models). We consider a few different molecules: C (small number of
levels and transitions), SO (large number of levels and transitions) as
well as CO and HCO\(^+\) (intermediate). On a laptop with i7-7700HQ
cores running on Ubuntu 22.04, \texttt{pythonradex} computed the model
grid faster than \texttt{RADEX} by factors of approximately 1.5 (C), 6
(SO), 7 (CO) and 3 (HCO\(^+\)). Running the same test on the
Multi-wavelength Data Analysis System (MDAS) operated by the National
Astronomical Observatory of Japan (Rocky Linux 8.9 with AMD EPYC 7543
CPUs) resulted in an even larger performance advantage:
\texttt{pythonradex} calculated the grid faster by factors of 13 (C), 10
(SO), 13 (CO) and 12 (HCO\(^+\))\footnote{\texttt{Jadex}
  (\citeproc{ref-jadex}{Svoboda, 2022}) claims a performance advantage
  of a factor \textasciitilde110 over \texttt{RADEX}.}.

\section{\texorpdfstring{Additional differences between \texttt{RADEX}
and
\texttt{pythonradex}}{Additional differences between RADEX and pythonradex}}\label{additional-differences-between-radex-and-pythonradex}

\subsection{Output flux}\label{output-flux}

\texttt{RADEX} computes line fluxes based on a ``background subtracted''
intensity given by
\((B_\nu(T_\mathrm{ex})-I_\mathrm{bg})(1-e^{-\tau})\), where \(B_\nu\)
is the Planck function, \(T_\mathrm{ex}\) the excitation temperature,
\(I_\mathrm{bg}\) the external background and \(\tau\) the
(frequency-dependent) optical depth. This may or may not be the right
quantity to be compared to observations. For example, it is not
appropriate when considering data from interferometers like ALMA.
\texttt{pythonradex} does not apply any observational correction, giving
the user the flexibility to decide how to compare the computed fluxes to
observations.

\subsection{Flux for spherical
geometry}\label{flux-for-spherical-geometry}

\begin{figure}
\centering
\includegraphics[keepaspectratio]{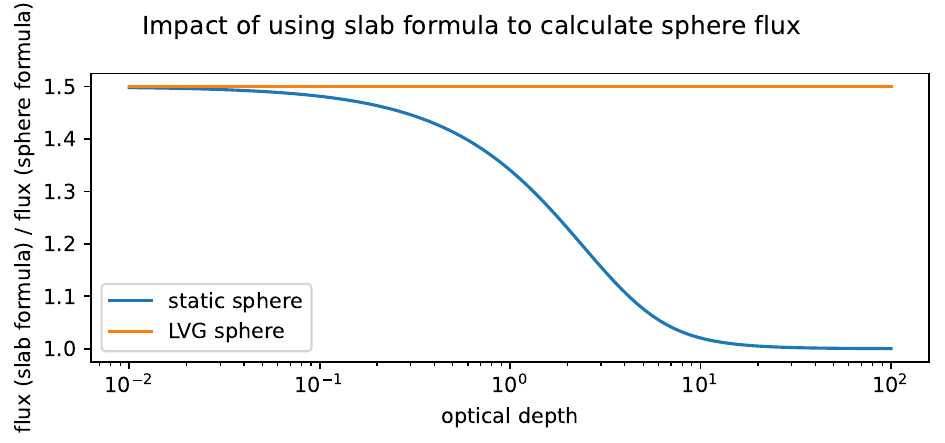}
\caption{For the two spherical geometries, fluxes using the slab formula
(as done by \texttt{RADEX}) and formulae appropriate for a sphere (as
done by \texttt{pythonradex}) were calculated. The figure shows the
ratio of the fluxes as function of optical
depth.\label{fig:flux_comparison_sphere}}
\end{figure}

To calculate line fluxes (energy per time per area, e.g.\ W/m$^2$), \texttt{pythonradex} uses different formulae
depending on the geometry (see the
\href{https://pythonradex.readthedocs.io/en/latest/index.html}{\texttt{pythonradex}
documentation} for more details). On the other hand, \texttt{RADEX}
always uses the formula for a slab. \autoref{fig:flux_comparison_sphere}
illustrates the consequences. For a static sphere, by using the slab
formula, the flux is overestimated by a factor 1.5\footnote{The factor
  1.5 corresponds to the volume ratio of a ``spherical slab'' (i.e.~a
  cylinder) to a sphere.} in the optically thin limit. In the optically
thick case, only the surface of the static sphere is visible, so the
different formulae give the same result. On the other hand, for the LVG
sphere, the difference is always a factor 1.5 regardless of optical
depth. This is a consequence of the LVG assumption that photons always
escape unless absorbed locally. By computing the optically thin flux
directly\footnote{In units of W/m\(^2\),
  \(F_\mathrm{thin} = V_\mathrm{sphere}n_2A_{21}\Delta E \frac{1}{4\pi d^2}\)
  with \(V_\mathrm{sphere}=\frac{4}{3}R^3\pi\) the volume of the sphere,
  \(n\) the number density, \(x_2\) the fractional level population of
  the upper level, \(A_{21}\) the Einstein coefficient, \(\Delta E\) the
  energy of the transition and \(d\) the distance.}, it can be confirmed
that the formulae used by \texttt{pythonradex} are correct.

\section{Dependencies}\label{dependencies}

\texttt{pythonradex} depends on the following packages:

\begin{itemize}
\tightlist
\item
  NumPy (\citeproc{ref-Harris2020}{Harris et al., 2020})
\item
  SciPy (\citeproc{ref-Virtanen2020}{Virtanen et al., 2020})
\item
  Numba (\citeproc{ref-Lam2015}{Lam et al., 2015})
\end{itemize}

\section{Acknowledgements}\label{acknowledgements}

I thank \href{https://github.com/andresmegias}{@andresmegias} and
\href{https://github.com/GijsVermarien}{@GijsVermarien} for their
thorough review which led to important improvements of the code and
documentation. I would also like to thank Simon Bruderer for his helpful
clarifications about the ALI method, Andrés Asensio Ramos for helpful
discussions about the LVG approximation and the \texttt{MOLPOP-CEP}
code, and \href{https://github.com/pscicluna}{@pscicluna} for helpful
comments on the code and paper draft. Performance testing was in part
carried out on the Multi-wavelength Data Analysis System operated by the
Astronomy Data Center (ADC), National Astronomical Observatory of Japan.

\section*{References}\label{references}
\addcontentsline{toc}{section}{References}

\protect\phantomsection\label{refs}


\begin{CSLReferences}{1}{0}
\bibitem[\citeproctext]{ref-AsensioRamos2018}
Asensio Ramos, A., \& Elitzur, M. (2018). {MOLPOP-CEP}: An exact, fast
code for multi-level systems. \emph{Astronomy and Astrophysics},
\emph{616}, A131. \url{https://doi.org/10.1051/0004-6361/201731943}

\bibitem[\citeproctext]{ref-Brinch2010}
Brinch, C., \& Hogerheijde, M. R. (2010). {LIME - a flexible, non-LTE
line excitation and radiation transfer method for millimeter and
far-infrared wavelengths}. \emph{Astronomy and Astrophysics},
\emph{523}, A25. \url{https://doi.org/10.1051/0004-6361/201015333}

\bibitem[\citeproctext]{ref-Dullemond2012}
Dullemond, C. P., Juhasz, A., Pohl, A., Sereshti, F., Shetty, R.,
Peters, T., Commercon, B., \& Flock, M. (2012). \emph{{RADMC-3D: A
multi-purpose radiative transfer tool}}. Astrophysics Source Code
Library, record ascl:1202.015.
\url{https://ui.adsabs.harvard.edu/abs/2012ascl.soft02015D}

\bibitem[\citeproctext]{ref-Elitzur1992}
Elitzur, M. (1992). Basic background concepts. In \emph{Astronomical
masers} (pp. 4--46). Springer Netherlands.
\url{https://doi.org/10.1007/978-94-011-2394-5_2}

\bibitem[\citeproctext]{ref-EMAA}
\emph{EMAA}. (2021). UGA, CNRS, CNRS-INSU, OSUG.
\url{https://doi.org/10.17178/EMAA}

\bibitem[\citeproctext]{ref-Harris2020}
Harris, C. R., Millman, K. J., Walt, S. J. van der, Gommers, R.,
Virtanen, P., Cournapeau, D., Wieser, E., Taylor, J., Berg, S., Smith,
N. J., Kern, R., Picus, M., Hoyer, S., Kerkwijk, M. H. van, Brett, M.,
Haldane, A., Río, J. F. del, Wiebe, M., Peterson, P., \ldots{} Oliphant,
T. E. (2020). Array programming with {NumPy}. \emph{Nature},
\emph{585}(7825), 357--362.
\url{https://doi.org/10.1038/s41586-020-2649-2}

\bibitem[\citeproctext]{ref-SpectralRadex}
Holdship, J., Vermariën, G., Keil, M., \& James, T. (2023).
SpectralRadex. In \emph{GitHub repository}. GitHub.
\url{https://github.com/uclchem/SpectralRadex}

\bibitem[\citeproctext]{ref-Lam2015}
Lam, S. K., Pitrou, A., \& Seibert, S. (2015). Numba: A LLVM-based
{Python} JIT compiler. \emph{Proceedings of the Second Workshop on the
LLVM Compiler Infrastructure in HPC}.
\url{https://doi.org/10.1145/2833157.2833162}

\bibitem[\citeproctext]{ref-Ng1974}
Ng, K.-C. (1974). {Hypernetted chain solutions for the classical
one-component plasma up to {\(\Gamma\)}=7000}. \emph{Journal of Chemical
Physics}, \emph{61}(7), 2680--2689.
\url{https://doi.org/10.1063/1.1682399}

\bibitem[\citeproctext]{ref-Rybicki1992}
Rybicki, G. B., \& Hummer, D. G. (1992). {An accelerated lambda
iteration method for multilevel radiative transfer. II. Overlapping
transitions with full continuum.} \emph{Astronomy and Astrophysics},
\emph{262}, 209--215.
\url{https://ui.adsabs.harvard.edu/abs/1992A&A...262..209R}

\bibitem[\citeproctext]{ref-Rybicki1985}
Rybicki, G. B., \& Lightman, A. P. (1985). Fundamentals of radiative
transfer. In \emph{Radiative processes in astrophysics} (pp. 1--50).
John Wiley \& Sons, Ltd. \url{https://doi.org/10.1002/9783527618170.ch1}

\bibitem[\citeproctext]{ref-Schoier2005}
Schöier, F. L., van der Tak, F. F. S., van Dishoeck, E. F., \& Black, J.
H. (2005). {An atomic and molecular database for analysis of
submillimetre line observations}. \emph{Astronomy and Astrophysics},
\emph{432}(1), 369--379.
\url{https://doi.org/10.1051/0004-6361:20041729}

\bibitem[\citeproctext]{ref-jadex}
Svoboda, B. (2022). Jadex. In \emph{GitHub repository}. GitHub.
\url{https://github.com/autocorr/Jadex.jl}

\bibitem[\citeproctext]{ref-ndradex}
Taniguchi, A. (2019). ndRADEX. In \emph{GitHub repository}. GitHub.
\url{https://github.com/astropenguin/ndradex}

\bibitem[\citeproctext]{ref-vanderTak2007}
van der Tak, F. F. S., Black, J. H., Schöier, F. L., Jansen, D. J., \&
van Dishoeck, E. F. (2007). {A computer program for fast non-LTE
analysis of interstellar line spectra. With diagnostic plots to
interpret observed line intensity ratios}. \emph{Astronomy and
Astrophysics}, \emph{468}(2), 627--635.
\url{https://doi.org/10.1051/0004-6361:20066820}

\bibitem[\citeproctext]{ref-Virtanen2020}
Virtanen, P., Gommers, R., Oliphant, T. E., Haberland, M., Reddy, T.,
Cournapeau, D., Burovski, E., Peterson, P., Weckesser, W., Bright, J.,
van der Walt, S. J., Brett, M., Wilson, J., Millman, K. J., Mayorov, N.,
Nelson, A. R. J., Jones, E., Kern, R., Larson, E., \ldots{} SciPy 1.0
Contributors. (2020). {SciPy} 1.0: Fundamental algorithms for scientific
computing in {Python}. \emph{Nature Methods}, \emph{17}, 261--272.
\url{https://doi.org/10.1038/s41592-019-0686-2}

\end{CSLReferences}
\end{document}